\documentclass[
    reprint,
    amsmath,
    amssymb,
    aps
]{revtex4-1}

\usepackage{float}
\usepackage{bbm}
\usepackage{epsfig}
\usepackage{epstopdf}
\usepackage{pgfplots}
\usepackage{amssymb}
\usepackage{amsmath}
\usepackage{scalefnt}
\usepackage{color}
\usepackage[title]{appendix}
\usepackage{hyperref}
\usepackage{qcircuit}
\usepackage{algorithm} 
\usepackage{algpseudocode}
\usepackage[capitalize]{cleveref}
\usepackage{braket,mleftright}
\usepackage{physics}
\usepackage{import}
\usepackage{dcolumn}
\usepackage{csquotes}
\usepackage{graphicx} 
\usepackage{subfigure}

\newcommand{\be}{\begin{equation}}
\newcommand{\ee}{\end{equation}}
\newcommand{\bea}{\begin{eqnarray}}
\newcommand{\eea}{\end{eqnarray}}
\def\ket#1{\left| #1 \right\rangle}

\begin{document}
\title{Designing globally optimal entangling gates using geometric space curves}

\author{Ho Lun Tang}
\affiliation{Department of Physics, Virginia Tech, Blacksburg, VA 24061, USA}
\author{Kyle Connelly}
\affiliation{Department of Physics, Virginia Tech, Blacksburg, VA 24061, USA}
\author{Ada Warren}
\affiliation{Department of Physics, Virginia Tech, Blacksburg, VA 24061, USA}
\author{Fei Zhuang}
\affiliation{Department of Physics, Virginia Tech, Blacksburg, VA 24061, USA}
\author{Sophia E. Economou}
\affiliation{Department of Physics, Virginia Tech, Blacksburg, VA 24061, USA}
\author{Edwin Barnes}
\affiliation{Department of Physics, Virginia Tech, Blacksburg, VA 24061, USA}

\date{\today}

\begin{abstract}
High-fidelity entangling gates are essential for quantum computation. 
Currently, most approaches to designing such gates are based either on simple, analytical pulse waveforms or on ones obtained from numerical optimization techniques. In both cases, it is typically not possible to obtain a global understanding of the space of waveforms that generate a target gate operation, making it challenging to design globally optimal gates. Here, we show that in the case of weakly coupled qubits, it is possible to find all pulses that implement a target entangling gate. We do this by mapping quantum evolution onto geometric space curves. We derive the minimal conditions these curves must satisfy in order to guarantee a gate with a desired entangling power is implemented. Pulse waveforms are extracted from the curvatures of these curves. We illustrate our method by designing fast, CNOT-equivalent entangling gates for silicon quantum dot spin qubits with fidelities exceeding 99\%. We show that fidelities can be further improved while maintaining low bandwidth requirements by using geometrically derived pulses as initial guesses in numerical optimization routines.
\end{abstract}

\maketitle
\section{Introduction}

High-fidelity entangling gates are a key requirement in all circuit-based approaches to quantum computing. Such gates are often implemented using electromagnetic pulse waveforms based on simple analytical shapes such as square or Gaussian functions, with fidelities optimized by adjusting amplitudes, timings, or bandwidths~\cite{PhysRevLett.103.110501, PhysRevA.83.012308, PhysRevA.100.022332,CalderonVargas_PRB2019, Magesan_PRA2020,PhysRevA.97.042348}. Numerical pulse-shape optimization recipes such as GRAPE are also commonly employed to improve fidelities further~\cite{KHANEJA2005296, PhysRevA.84.022326, PhysRevA.97.042348,PhysRevB.96.024504, Yang2019}. Such methods have led to experimental demonstrations of high-fidelity entangling gates in silicon quantum dots~\cite{Veldhorst_Nature2015,Huang2019,xue_computing_2021, noiri_fast_2021, mills_two-qubit_2021, Leon2021, PRXQuantum.2.010303}, superconducting qubits~\cite{Barends_Nature2014,Sheldon_PRA2016,Krantz_APR2019,Rol_PRL2019,Gao_Nature2019},  and trapped ions~\cite{Egan2021, Srinivas2021, PRXQuantum.2.020343}.

Despite the substantial progress that has been made in recent years, further improvements in entangling gates are still widely needed. In addition to the very high fidelities that are required to comfortably exceed error correction thresholds~\cite{PhysRevA.83.020302,PhysRevA.86.032324}, it is also important to reduce gate times and pulse amplitude and bandwidth requirements as much as possible to speed up algorithms and lessen the technological overhead. Finding pulse waveforms that are optimal with respect to all these factors is generally a challenging task. While numerical techniques address this issue to a large extent, it is often still difficult to obtain globally optimal solutions with numerical methods alone. This motivates the search for complementary methods that provide a global, analytical understanding of the space of pulse waveforms that generate a target gate operation. Such methods, in combination with numerical optimization protocols, could lead to significant enhancements in gate performance.

Recently, it has been shown that there exists a correspondence between quantum evolution and geometric space curves \cite{Barnes_2022}. This connection provides a global perspective on the relationship between pulses and the quantum evolution they generate.
This has been exploited to design single- and two-qubit gates that dynamically correct for noise that acts transversely to the driving field \cite{Zeng_2018,PhysRevA.98.012301,PhysRevA.99.052321,PRXQuantum.2.010341}. In this approach, noise-resistant pulse waveforms are obtained from the geometric properties of closed space curves.
This technique was also applied to study the speed limit of dynamically corrected gates \cite{PhysRevA.98.012301}, noise-resilient Landau-Zener transitions \cite{Zhuang2022noiseresistant}, and ``doubly geometric" gates that are simultaneously robust against pulse errors and transverse noise  \cite{PRXQuantum.2.030333}.

In this work, we show that entanglement growth in weakly coupled two-qubit systems can also be mapped to space curves. We use this mapping to develop a geometric framework for finding infinitely many pulses that generate a target entangling gate. 
The pulses that produce this entanglement can be obtained from the generalized curvatures of the curves. This provides a general method for finding pulses that implement gates with a target entangling power. We demonstrate this method in the context of silicon quantum dot spin qubits, where we derive the minimal space-curve conditions needed to produce CNOT-equivalent gates. We show that the pulses extracted from these curves generate maximally entangling gates with fidelities exceeding 99\% and gate times below 30 ns for typical experimental parameters. We show that using these geometrically engineered pulses as a seed in numerical optimization routines can further improve the fidelities beyond what is achievable with random initial guesses.

The paper is organized as follows. In Sec.~\ref{sec:system}, we introduce the two-qubit silicon spin system example that we focus on throughout this work. 
In Sec.~\ref{sec:geo}, we present our geometric framework that relates two-qubit entanglement growth to space curves in three dimensions.
We provide explicit examples of driving pulses and the corresponding high-fidelity entangling gates they generate.
In Sec.~\ref{sec:GRAPE}, we examine how the geometrically derived pulses perform after numerical optimization.

\section{Spin qubit System}\label{sec:system}

Throughout this work, we focus on the case of silicon quantum dot spin qubits to illustrate our approach, although the basic idea can be applied to any weakly coupled qubit system. Single-electron spin qubits in silicon quantum dots are a promising platform for quantum computation, due to long coherence times, the availability of all-electrical control, and the potential for scalability afforded by the existing silicon manufacturing infrastructure \cite{PhysRevA.57.120, RevModPhys.85.961, Gonzalez-Zalba_NatureElectron2021,Zwerver_NatureElectron2022}. 
While high-fidelity single- and two-qubit gates were demonstrated by several groups recently \cite{Yoneda2018, Huang2019, noiri_fast_2021, mills_two-qubit_2021,xue_computing_2021, noiri_fast_2021, mills_two-qubit_2021}, further improvements in two-qubit gate fidelities are still needed for most error correction schemes.
The fidelity is mainly limited by charge noise and nuclear spin bath fluctuations~\cite{Huang2018, PhysRevApplied.12.044054, PhysRevApplied.10.044017}.
Since the natural abundance of the spinful isotope $^{29}\text{Si}$ is only $4.7\%$, and can be further reduced by isotopic purification~\cite{Itoh2014}, charge noise is widely considered the dominant source of noise in the system.

The specific system we focus on consists of two electrons trapped in a silicon double quantum dot (DQD). We begin by deriving an effective Hamiltonian for this system following the analysis of Ref.~\cite{PhysRevB.100.035304}. In the following section, we use this effective Hamiltonian to establish the correspondence between two-qubit evolution and space curves.
The two electron spins are subject to an external magnetic field from a micromagnet, which is designed to maximize $dB_z/dx$ and $dB_y/dz$. The first term gives rise to the different Zeeman splittings of the two qubits since they are separated along the $x$-axis, well separating the resonance frequencies of the two qubits.
The strong gradient along the $z$-axis, i.e. $dB_y/dz$, is designed for electric dipole spin resonance (EDSR) control~\cite{doi:10.1126/science.1148092}.
By applying a microwave pulse to the metal gate, we can oscillate the position of the electrons primarily in the $z$-axis, which leads to an oscillating $B_y$ magnetic field, enabling EDSR control.
The exchange interaction between the two spins, $J(t)$, can be tuned by changing the energy barrier via a middle metal gate which also controls the separation between the dots~\cite{Martins_PRL2016,Reed_PRL2016}.
Our starting point is the Heisenberg Hamiltonian,
\begin{equation}
\label{eq:Heisen}
    H(t) = J(t)\left(S_L\cdot S_R-\frac{1}{4}\right) + S_L\cdot B_L+S_R\cdot B_R,
\end{equation}
where $S_{L}$ ($S_R$) is the spin operator of the electron in the left (right) quantum dot. 
The external magnetic field $B_{L,R}$ has two components along the $y$ and $z$-axes.
The $y$-component is time-dependent due to the drive field causing the electron oscillation in $z$-direction, i.e., $B_{y,q}=B^0_{y,q}+B^1_{y,q}(t)\cos\left(\omega t +\phi\right)$ , where $q=L/R$, $\omega$ and $\phi$ being the driving frequency and phase, respectively.
On the other hand, the $z$-component of the magnetic field, which sets the Zeeman splitting of the two qubits, is kept constant throughout the control process.
It is only slowly changed while the exchange coupling $J$ is being turned on and off adiabatically, but in general, it is a function of the exchange coupling $B_{z,q}(J)=B^0_{z,q}+B^1_{z,q}(J)$, where the first term is the field when $J=0$.

Following Ref.~\cite{PhysRevB.100.035304}, we work with the eigenbasis set of the undriven version of the Hamiltonian in Eq.~\eqref{eq:Heisen}, i.e. $B_{y,q}=0$, where the computational basis $\ket{\uparrow \downarrow}$ and $\ket{\downarrow \uparrow}$ are superposed into $\widetilde{\ket{\uparrow \downarrow}}$ and $\widetilde{\ket{\downarrow \uparrow}}$ by the exchange coupling.
We then go to the interaction picture defined by the undriven and uncoupled Hamiltonian ($B_{y,q}=J=0$).
Together with the approximation justified by the larger Zeeman splitting deference between the two qubits compared to the exchange coupling, i.e. $J\ll |B_{z,L}-B_{z,R}|$, the interaction picture Hamiltonian becomes
\begin{widetext}
\begin{equation}
    H_{\mathrm{int}}=\frac{1}{2}
    \begin{pmatrix}
    2\Bar{B}^1_z & -i(B_{y,L}+\xi B_{y,R})e^{-i\alpha_-t} & -i(B_{y,R}-\xi B_{y,L})e^{i\alpha_+t} & 0 \\
    i(B_{y,L}+\xi B_{y,R})e^{i\alpha_-t} & \Delta B_z^1-J + \frac{J\xi}{2} & 0 & -i(B_{y,R}+\xi B_{y,L})e^{i\alpha_+t} \\
    i(B_{y,R}-\xi B_{y,L})e^{-i\alpha_+t} & 0 & -\Delta B_z^1-J - \frac{J\xi}{2} & -i(B_{y,L}-\xi B_{y,R})e^{-i\alpha_-t} \\
    0 & i(B_{y,R}+\xi B_{y,L})e^{-i\alpha_+t} & i(B_{y,L}-\xi B_{y,R})e^{i\alpha_-t} & -2\Bar{B}^1_z
    \end{pmatrix},
    \label{eq:full}
\end{equation}
\end{widetext}
where $\Bar{B}_z=\left(B_{z,L}^0+B_{z,R}^0\right)/2$, $\Bar{B}_z^1=\left(B_{z,L}^1+B_{z,R}^1\right)/2$, $\Delta B_z= B_{z,R}^0-B_{z,L}^0$, $\Delta B_z^1= B_{z,R}^1-B_{z,L}^1$, $\alpha_\pm=(\Delta B_z\pm2\Bar{B}_z)/2$, and $\xi=J/(\Delta B_z+\Delta B_z^1)$. 
This is the Hamiltonian used in the numerical calculations throughout this work, and we use the parameters reported in Ref. \cite{doi:10.1126/science.aao5965}: $B_{z,L}^0/2\pi=18.287$ GHz, $B_{z,R}^0/2\pi=18.501$ GHz, $B_{z,L}^1/2\pi=52.71$ MHz, $B_{z,R}^1/2\pi=5.76$ MHz, $B_{y,L}^0/2\pi=5$ MHz, $B_{y,R}^0/2\pi=55$ MHz and $\phi=3\pi/2$.

To gain a better understanding of the evolution generated by this Hamiltonian, it helps to further simplify the model by switching to a rotating frame, $H_{\mathrm{rot}}= U_\omega H U_\omega^\dagger-iU_\omega \Dot{U}_\omega^\dagger$, with $U_\omega=\exp\left[i\omega t\left(S_{z,R}+S{z,L}/\hbar\right)\right]$, and applying the rotating wave approximation. 
Since our target is to obtain a certain entangling gate up to local unitaries and the rotating frame evolution operator is only different from the interaction frame by local unitaries, the gate in both frames has the same entangling properties.
By matching the driving frequency $\omega$ to the resonance frequency of the left qubit, the driving on the right qubit becomes negligible compared to the Zeeman splitting difference, $B^1_{y,R}\ll \Delta B_z+\Delta B_z^1$.
Ultimately, we obtain the simplified Hamiltonian:
\begin{align}
    H_{\mathrm{rot}}=&\ \frac{1}{4}
    \begin{pmatrix}
    0 & B^1_{y,L}(t) & 0 & 0 \\
    B^1_{y,L}(t) & 0 & 0 & 0 \\
    0 & 0 & -2J & B^1_{y,L}(t)\\
    0 & 0 & B^1_{y,L}(t) & 2J
    \end{pmatrix}\\
    =&\ \frac{J}{4}(ZZ-IZ)+\frac{B_{y,L}^1}{4}IX.
    \label{eq:sim}
\end{align}
This describes two qubits coupled by a weak Ising interaction, and only the second qubit is driven.

\section{Geometric space curves and corresponding pulses}\label{sec:geo}
In previous works that used geometric space curves to design pulses \cite{Zeng_2018, Zhuang2022noiseresistant, PhysRevA.99.052321, PhysRevA.98.012301, PRXQuantum.2.010341, PRXQuantum.2.030333, Barnes_2022}, the space curves represented the effect of noise errors on the quantum evolution. Specifically, the net displacement between the initial and final points of the curve quantified the importance of the first-order term in a perturbative expansion of the evolution operator in powers of the noise error. In this work, we also use geometric space curves, but instead of quantifying the error due to noise, the space curve here represents the entanglement generated during the evolution.

Following the Hamiltonian in Eq.~\eqref{eq:sim}, we define the single-qubit terms of the Hamiltonian as $H_0=-\frac{J}{4}IZ+\frac{\Omega(t)}{4}IX$, where $\Omega(t)= B_{y,L}^1$ is the driving pulse. 
We then switch to the rotating frame defined by $H_0$ so that the two-qubit interaction term $\frac{J}{4}ZZ$ is isolated. Again this rotating frame transformation only involves local unitaries, and so has no effect on the entangling properties.
The evolution operator in this frame has the form
    \begin{align}
   \tilde{U} =&\ \mathcal{T}\exp\left[-i\frac{J}{4}\int dt U_0^\dagger ZZ U_0\right]\nonumber \\
     \approx&\ \exp\left[-i\frac{J}{4}\int dt U_0^\dagger ZZ U_0\right]\nonumber \\
    = &\ \exp\left[-i\frac{J}{4}\left(R_1(t) ZX+R_2(t) ZY + R_3(t) ZZ\right)\right].
    \label{eq:magnus}
\end{align}
Here we have kept only the first-order term in the Magnus expansion of the evolution operator under the assumption that the coupling is weak.
The three components \{$R_1(t), R_2(t), R_3(t)$\} define the coordinates of a 3D space curve $\vec{R}(t)$ parameterized by evolution time $t$, i.e.,
\begin{equation}
\label{eq:util}
    \tilde{U} \approx\ \exp\left[-i\frac{J}{4}Z\otimes \left(\Vec{R}(t)\cdot \Vec{\sigma}\right)\right].
\end{equation}

\subsection{Makhlin Invariants}
In order to study the entangling properties of this two-qubit gate, we consider the Makhlin invariants introduced in Ref.~\cite{Makhlin2002}:
\begin{align}
    G_1&=\ \frac{[\text{Tr} (M)]^2}{16\det U}, \\
    G_2&=\ \frac{[\text{Tr} (M)]^2-\text{Tr} (M^2)}{4\det U},
\end{align}
where $U$ is a two-qubit unitary, and $M$ is its symmetrized version expressed in the Bell basis:
\begin{align}
    M&=\ \left(Q^\dagger U Q\right)^T\left(Q^\dagger U Q\right), \\
    Q &= \frac{1}{\sqrt{2}}\begin{pmatrix}
    1 & 0 & 0 & i \\
    0 & i & 1 & 0 \\
    0 & i & -1 & 0 \\
    1 & 0 & 0 & -i
    \end{pmatrix}.
\end{align}
These two quantities are invariant under local operations so that they encode only the entanglement information, and any two-qubit unitaries with the same Makhlin invariants are equivalent up to local operations.
Here, we aim to design two-qubit entangling gates with specified Makhlin invariants.

Using the fact that $QQ^T=-YY$, the trace of $M$ can be rewritten as
\begin{equation}
    \text{Tr} (M) = \text{Tr}\left(YY U^T YY U\right).
\end{equation}
We further exploit the fact that the symmetric two-qubit Pauli terms in Eq.~\eqref{eq:util} commute with $YY$, while the anti-symmetric term anticommutes with $YY$, therefore $YY\tilde{U}^T YY=\tilde{U}$.
The trace of $M$ can be further simplified
\begin{align}
    \text{Tr}\left(M\right) &= \text{Tr}\left(\tilde{U}^2\right) = 4\cos \left(\frac{J}{2}|\Vec{R}|\right), \\
    \text{Tr}\left(M^2\right) &= \text{Tr}\left(\tilde{U}^4\right) = 4\cos \left(J|\Vec{R}|\right).
\end{align}
The corresponding Makhlin invariants read as
\begin{align}
    G_1&=\cos^2\left(\frac{J}{2}|\vec{R}|\right), \\
    G_2&=2+\cos\left(J|\vec{R}|\right).
\end{align}
We can see that at $t=0$, $\vec{R}(0)=0$, and hence $G_1=1$ and $G_2=3$, which correspond to the identity operation.
When $J|\vec{R}|=(2n+1)\pi$, we have $G_1=0$ and $G_2=1$, which correspond to a CNOT-equivalent gate.
Generally the Makhlin invariants for a conditional $X$-rotation $R_X(\theta)$ are 
\begin{align}
    G_1&=\cos^2\left(\theta\right), \\
    G_2&=2+\cos\left(2\theta\right),
\end{align}
and thus one can obtain a controlled-$R_X(\theta)$ for any arbitrary angle by tuning the final displacement $|\vec{R}(t_f)|$.

\subsection{Geometric properties of the space curve}
After designing the 3D space curve $\vec{R}(t)$ with the desired final displacement, the corresponding driving pulse can be read from the geometric properties of the curve.
The local curvature $\kappa(t)$ and torsion $\tau(t)$ can be obtained from time-derivatives of $\vec{R}(t)$:
\begin{align}
    \kappa_R =&\ \left|\ddot{\vec{R}}\right| = \frac{\Omega}{2}, \label{eq:kappaR}\\
    \tau_R =&\ \frac{\left(\dot{\vec{R}}\times \ddot{\vec{R}}\right)\cdot \dddot{\vec{R}}}{\left|\dot{\vec{R}}\times \ddot{\vec{R}}\right|^2}=\frac{J}{2}.
\end{align}
We see that the driving pulse is equal to twice the curvature of the space curve, while the torsion is  $J/2$. This means that all constant-torsion curves $\vec{R}(t)$ with $\tau_R=J/2$ and final displacement $|\vec{R}(t_f)|$ yield a pulse $\Omega(t)=2\kappa_R$ that generates a controlled--$R_X(\theta)$ gate with $\theta=J|\vec{R}(t_f)|/2$.
To design a constant-torsion curve, we follow the approach of Ref.~\cite{Zhuang2022noiseresistant}. Consider the coordinate system defined by the three orthonormal vectors \{$\hat{T}$, $\hat{N}$, $\hat{B}$\} associated with the curve, where  $\hat{T} = \dot{\vec{R}}$ is the tangent vector, $\hat{N}=\dot{ \hat{T}}/\left|\dot{ \hat{T}}\right|$ is the normal vector, and $\hat{B}$ is the binormal vector, given by $\hat{B}=\hat{T}\times\hat{N}$. These vectors obey the Frenet-Serret equations,
\begin{equation}
    \dv{t}\mqty(\hat{T} \\ \hat{N} \\ \hat{B}) = 
    \mqty(  0 & \kappa_R & 0 \\
            -\kappa_R & 0 & \tau_R \\
            0 & -\tau_R & 0
    ) \mqty(\hat{T} \\ \hat{N} \\ \hat{B}).
\end{equation}
By manipulating the relation between \{$\hat{T}$, $\hat{N}$, $\hat{B}$\}, we find
\begin{align}\label{eq:RfromB}
    \vec{R}(t)=& \int_0^t \hat{T}(t') d t' 
    = \int_0^t \hat{N}(t')\times \hat{B}(t') dt'\nonumber\\
    =& -\int_0^t \frac{1}{\tau_R}\frac{d\hat{B}}{dt'}\times \hat{B}(t') dt' = \frac{1}{\tau_R}\int \hat{B}\times d \hat{B}.
\end{align}
We can interpret $\hat{B}(t)$ as a curve in its own right. Thus, we see that starting from an arbitrary $\hat{B}(t)$ curve, which lies on a unit sphere, we can use this formula to find a space curve with constant torsion.
Note that since $\abs{\dot{\hat{B}}} = \abs{-\tau_R \hat{N}} = \tau_R$, the arclength along the binormal curve is given by $\tau_R t$, and the curvature of the binormal curve is given by $\kappa_B = \abs{\dv*[2]{\hat{B}}{(\tau_R t)}}$.

We would also like to find a driving pulse which starts and ends at zero amplitude, since these are typically easiest to implement in hardware. 
This property can be translated to the space curve side by rewriting the curvature in terms of the TNB frame of the $\hat{B}$ curve:
\begin{align}
    \kappa_R &= \abs{\dv{\hat{T}}{t}} = \abs{\dv{t}\qty(\frac{1}{\tau_R} \hat{B} \times \dv{\hat{B}}{t})} 
    = \tau_R \abs{\hat{B} \times \dv[2]{\hat{B}}{(\tau_R t)}} \nonumber \\
    &= \tau_R \kappa_B \abs{\hat{B} \times \hat{N}_B} = \tau_R \kappa_B \sin\theta.
\end{align}
Here $\theta$ denotes the angle between $\hat{B}$ and $\hat{N}_B$. Since $\hat{B}$ is a normal vector of the sphere, the term $\kappa_B \sin\theta$ can be recognized as the geodesic curvature of the binormal curve, $\kappa_{B,g}$, and the expression above becomes $\kappa_R/\tau_R = \kappa_{B,g}$. Thus in order for the pulse amplitude to start and end at zero, $\kappa_{B,g}$ must vanish at $t=0$ and $t=t_f$. This is equivalent to requiring that $\hat{B}$ be parallel to $\hat{N}_B$ at $t=0$ and $t=t_f$, or equivalently that $\hat{B}$ trace a great circle in infinitesimal neighborhoods around $t=0$ and $t=t_f$.

\subsection{Examples}
To demonstrate the above method, we now show examples of space curves satisfying the conditions described above and their corresponding pulses.
We use the following ansatz for the $\hat{B}$ curve:
\begin{equation}
    \sqrt{1-\lambda \sin ^2 \beta l} \left(\cos l, \sin l, 0\right)+\sqrt{\lambda}\sin \beta l \left(0, 0, 1\right),
    \label{eq:b_ansatz}
\end{equation}
where $\lambda$ and $\beta$ are parameters that can be tuned to achieve a desired value of $J|\vec{R}(t_f)|$, and $l$ parameterizes the $\hat{B}$ curve, ranging from $0$ to $l_f=\pi/\beta$. 
Here we choose specific values of $\beta$ and perform a linear search for $\lambda\in\left[0,1\right]$ to achieve $J|\vec{R}(t_f)|=(2n+1)\pi$.
This ansatz starts and ends as a great circle which ensures that the corresponding pulse starts and ends at zero.

\begin{figure*}%
\subfigure{%
  \includegraphics[width=0.45\textwidth]{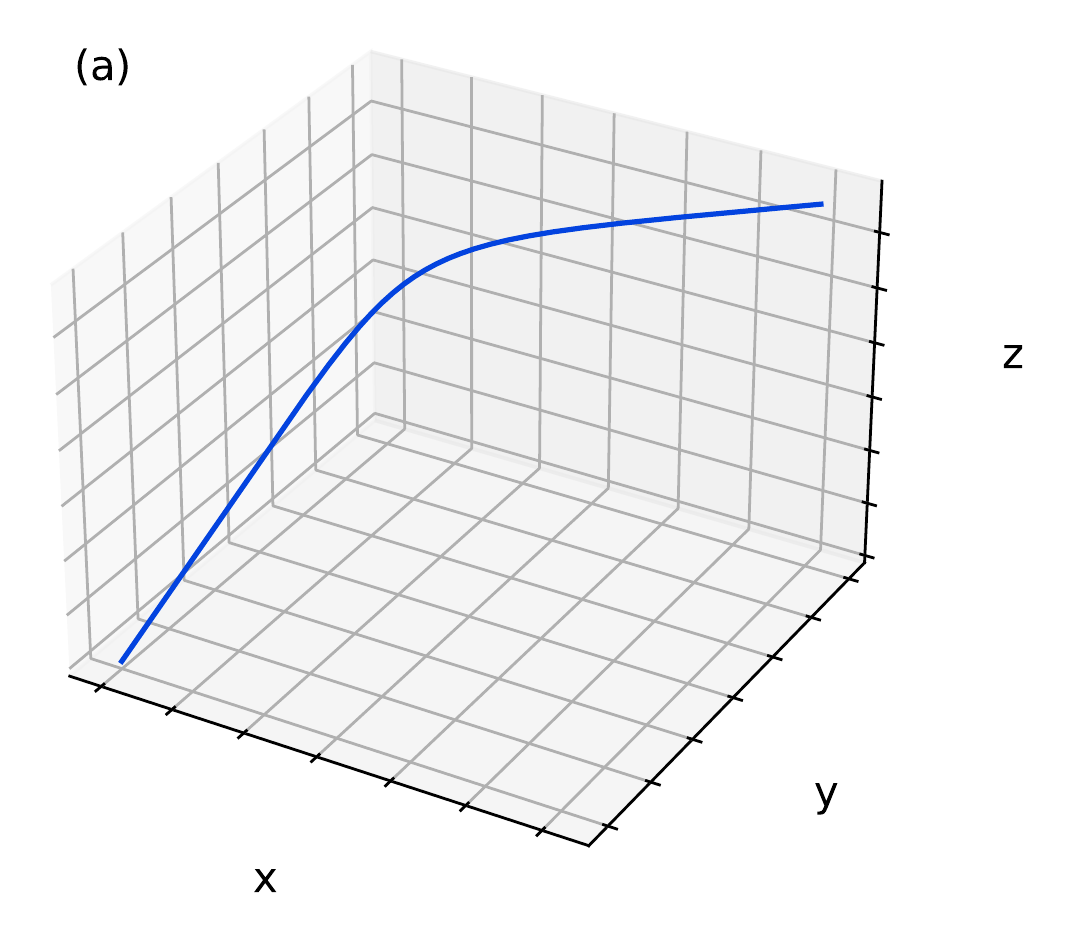}%
}%
\hspace*{\fill}
\subfigure{
  \includegraphics[width=0.45\textwidth]{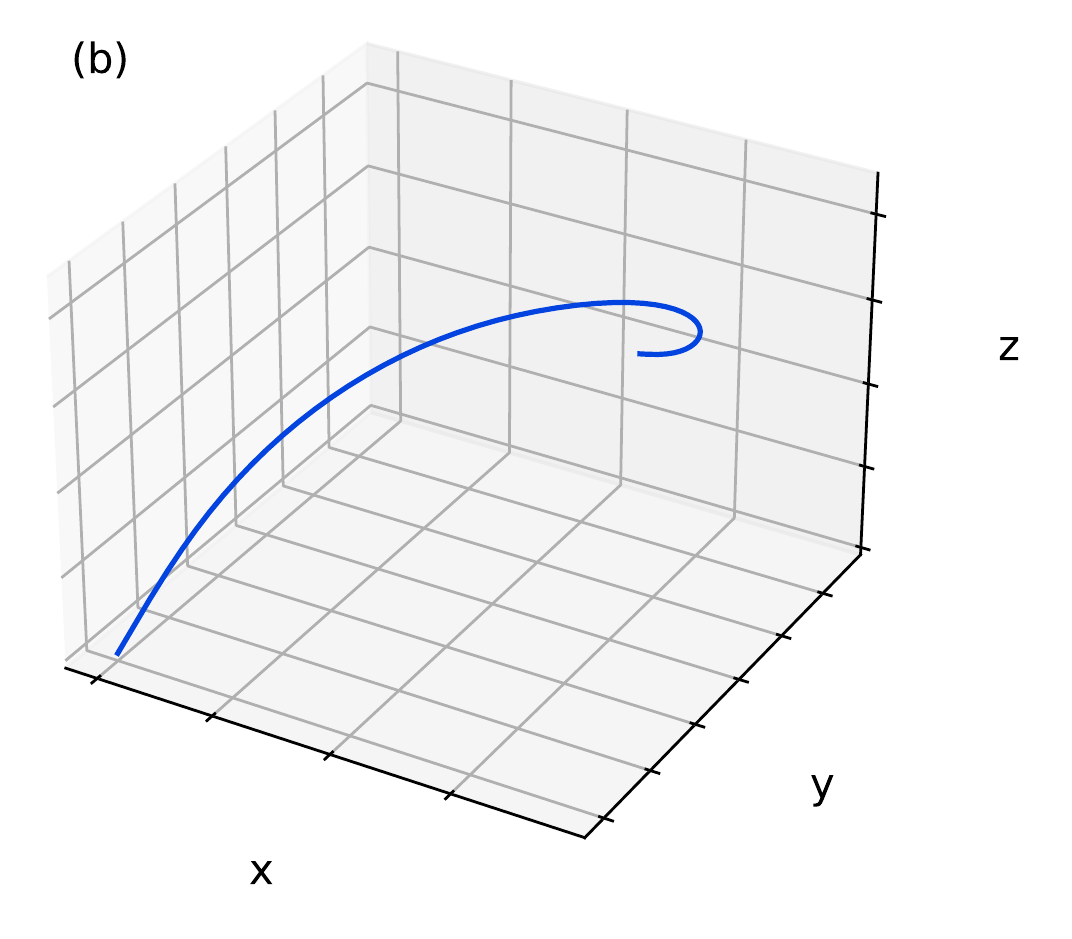}%
}%

\subfigure{%
  \includegraphics[width=0.45\textwidth]{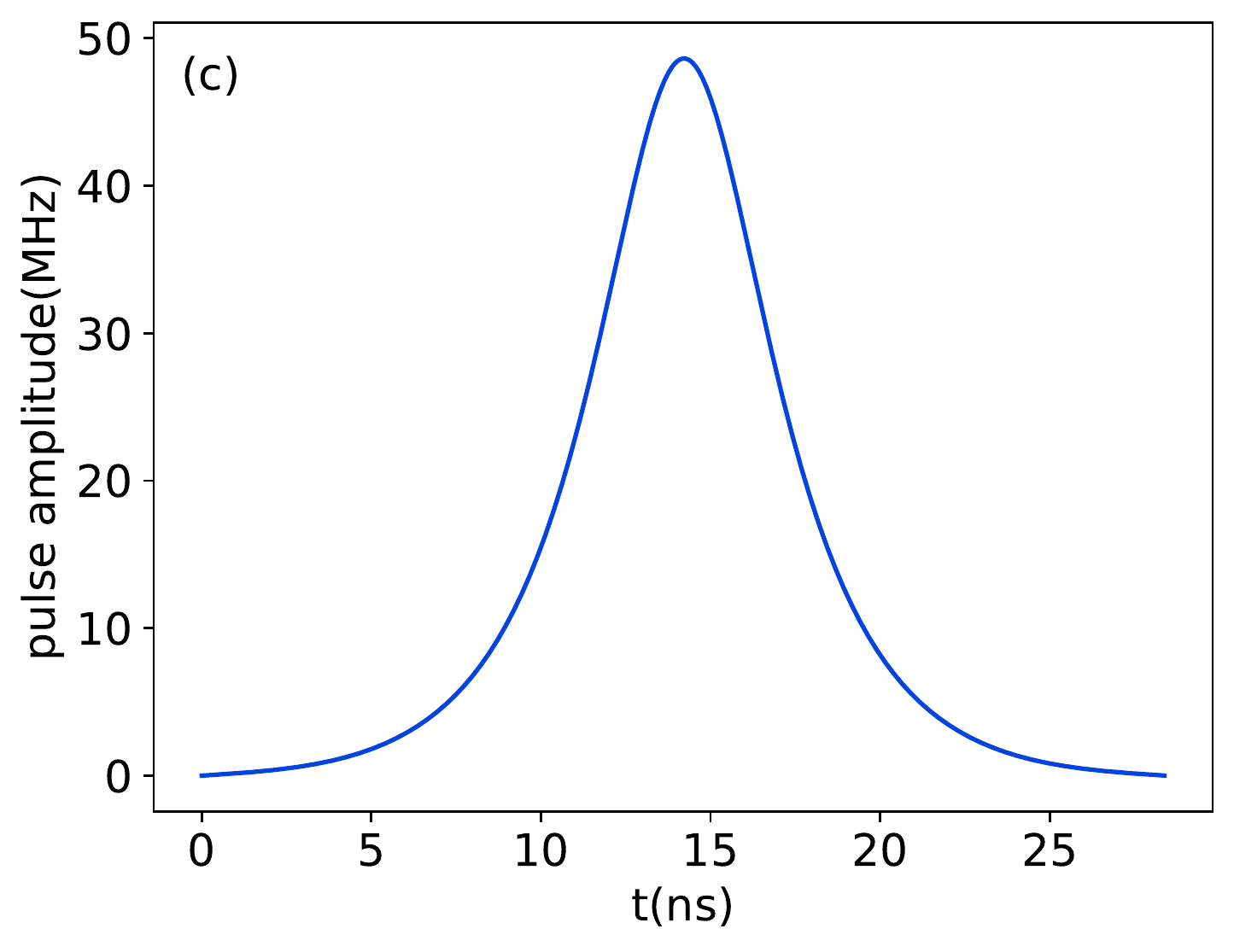}%
}%
\hspace*{\fill}
\subfigure{
  \includegraphics[width=0.45\textwidth]{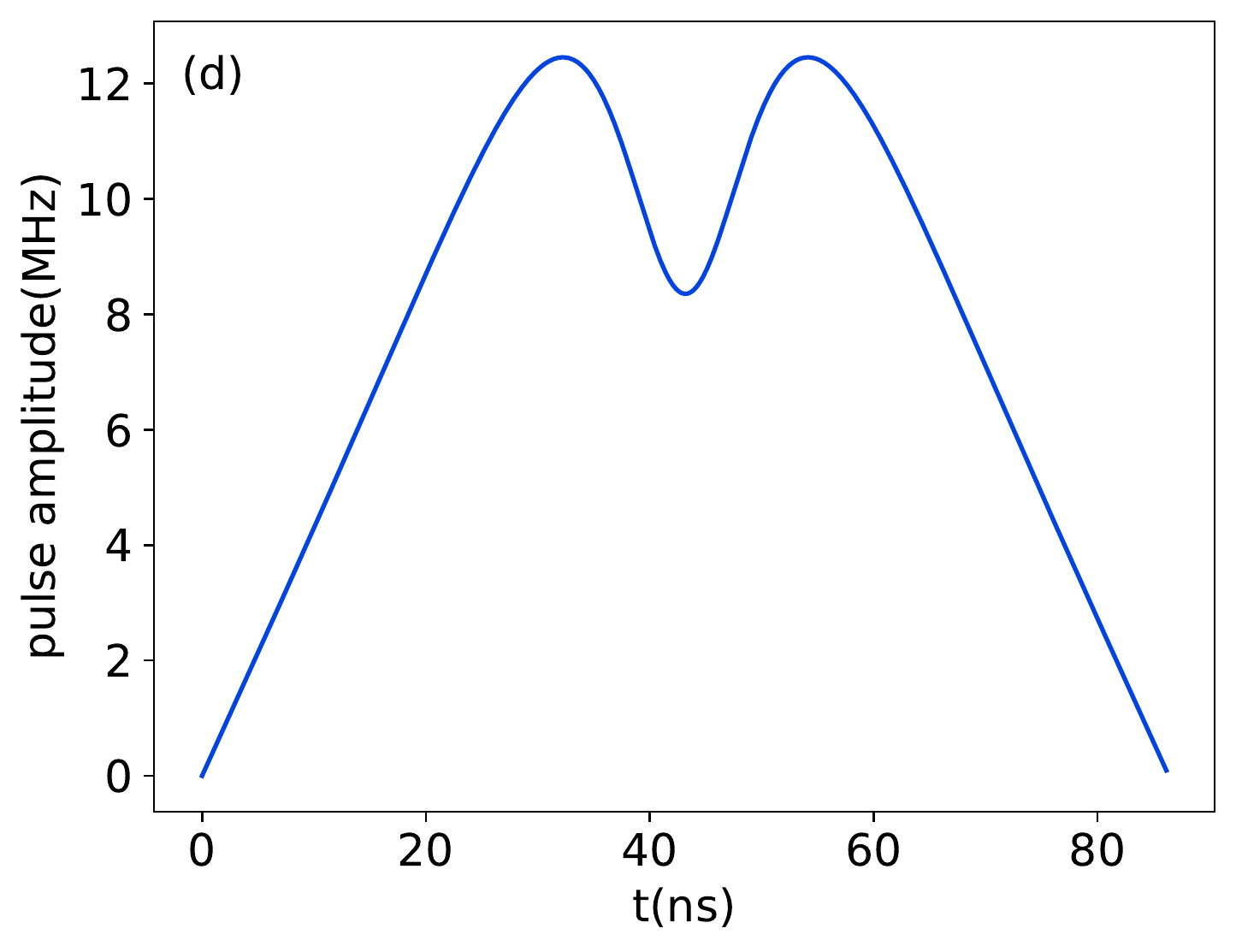}%
}%

\caption{(a) 3D space curve with $J|\vec{R}(t_f)|=\pi$, found by setting $\beta=2\pi/3$ and $\lambda=0.221163$. (b) 3D space curve with $J|\vec{R}(t_f)|=3\pi$, found by setting $\beta=\pi/6$ and $\lambda=0.561651$. (c) Corresponding driving pulse of (a) starts and ends at zero. It takes $28.3836$ ns with fidelity $99.84\%$ by complementing the two-qubit gate with 4 single-qubit gates. (d) Corresponding driving pulse of curve in (b) with gate time $86.2373$ ns and fidelity $99.43\%$.}\label{fig:curve}
\end{figure*}

We obtain the corresponding pulse by first computing the space curve $\vec{R}(t)$ from $\hat{B}$ using Eq.~\eqref{eq:RfromB} and then employing Eq.~\eqref{eq:kappaR}. We then numerically solve the Schr\"{o}dinger equation with the full Hamiltonian in Eq.~\eqref{eq:full} to obtain the evolution operator $U_{\mathrm{int}}$.
Since the driving pulses are designed to create gates locally equivalent to a CNOT gate, we apply local unitaries before and after $U_{\mathrm{int}}$ to bring it as close as possible to a CNOT:
\begin{equation}
    U = K_1 U_{\mathrm{int}} K_2,
\end{equation}
where the $K_i$ are tensor products of single-qubit gates on both qubits, so each $K_i$ depends on 6 rotation angles.
The fidelity of $U$ with a CNOT is then calculated using the formula
\begin{equation}
    F=\frac{1}{n(n+1)}\left[\text{Tr}(U^\dagger U)+\abs{\text{Tr}(U_{\mathrm{targ}}^\dagger U)}^2\right],
    \label{eq:fid}
\end{equation}
where $n$ is the Hilbert space dimension, and $U_{\mathrm{targ}}$ is the target gate (i.e. a CNOT gate). The local unitaries $K_i$ are chosen to maximize $F$.
Fig.~\ref{fig:curve} shows two examples of curves that yield CNOT gates up to local unitaries. Panels (a) and (b) show space curves $\vec{R}(t)$ generated from the $\hat{B}$ ansatz Eq.~\ref{eq:b_ansatz}, with $J\abs{\vec{R}(t_f)} = \pi$ and $J\abs{\vec{R}(t_f)} = 3\pi$, respectively. Panels (c) and (d) then show the pulses derived from these space curves. These pulses achieve fidelities of $99.84\%$ and $99.43\%$, respectively.

This fidelity could be further increased by taking into account higher order terms in the Magnus expansion of $\tilde{U}$ in Eq.~\ref{eq:magnus}, which would give perturbative corrections to the driving pulse. However, it is much simpler to use our first-order pulses as initial guesses for numerical optimization.

\section{Space curve pulses as the initial guess for GRAPE}\label{sec:GRAPE}

Although some optimal control theory problems may be solved exactly, in many realistic cases one must resort to finding a numerical solution. In quantum control theory, one of the simplest and most popular numerical algorithms is GRadient Ascent Pulse Engineering, or GRAPE \cite{khaneja2005optimal}. GRAPE works by discretizing the control pulse and then using a gradient ascent algorithm to find a pulse that maximizes the fidelity of $U$ with some target gate $U_{\mathrm{targ}}$. Specifically, to control a system with Hamiltonian $H(t) \equiv H_0 + \Omega(t) H_c$, the duration $t_f$ of the pulse is broken up into $N$ subintervals of length $\Delta t = t_f/N$, and $\Omega(t)$ is taken to be piecewise constant within each subinterval, $\Omega(t) = \Sigma_{k=0}^{N-1} \Omega_k \Theta(t - k \Delta t)\Theta((k+1)\Delta t - t)$. $U$ is then approximated as 
\begin{align}
    U &= U_{N-1} U_{N-2} \cdots U_1 U_0,\nonumber \\ U_k &= \exp( -i \Delta t (H_0 + \Omega_k H_c) ).
\end{align}
GRAPE then treats the pulse amplitudes $\Omega_k$ as free parameters, and numerically searches for values that maximize the fidelity of $U$ with $U_{\mathrm{targ}}$ (Eq.~\eqref{eq:fid}). This can be done using a gradient ascent optimization, using the chain rule along with $\pdv{U}{\Omega_k} \approx -i \Delta t U_{N-1} \cdots U_{k+1} H_c U_k \cdots U_0$, valid to first order in $\Delta t$.
Here $U$ only needs to be locally equivalent to a CNOT, and so we modify GRAPE to minimize a cost function $C$ given by the difference between the Makhlin invariants of $U$ and CNOT:
\begin{equation}
C = \abs{G_1}^2 + \abs{G_2 - 1}^2.
\end{equation}

Although GRAPE is certainly a useful algorithm, it does have significant limitations. Like any gradient-based optimization algorithm, GRAPE will find a locally optimal solution, but it will typically not find a globally optimal solution if the initial pulse is not close to the global optimum. Additionally, in practice, the locally optimal pulses found by GRAPE often have undesirable properties for experimental implementation, such as being discontinuous, having high bandwidth, starting and/or ending at nonzero values, or changing signs. These issues can be circumvented by starting from a pulse obtained from the geometric formalism, since geometric pulses are analytically guaranteed to be near an optimal solution, and can be designed by hand to avoid the issues mentioned above.

\begin{figure}
    \centering
    \begin{subfigure}
        \centering
        \includegraphics[width=0.45\textwidth]{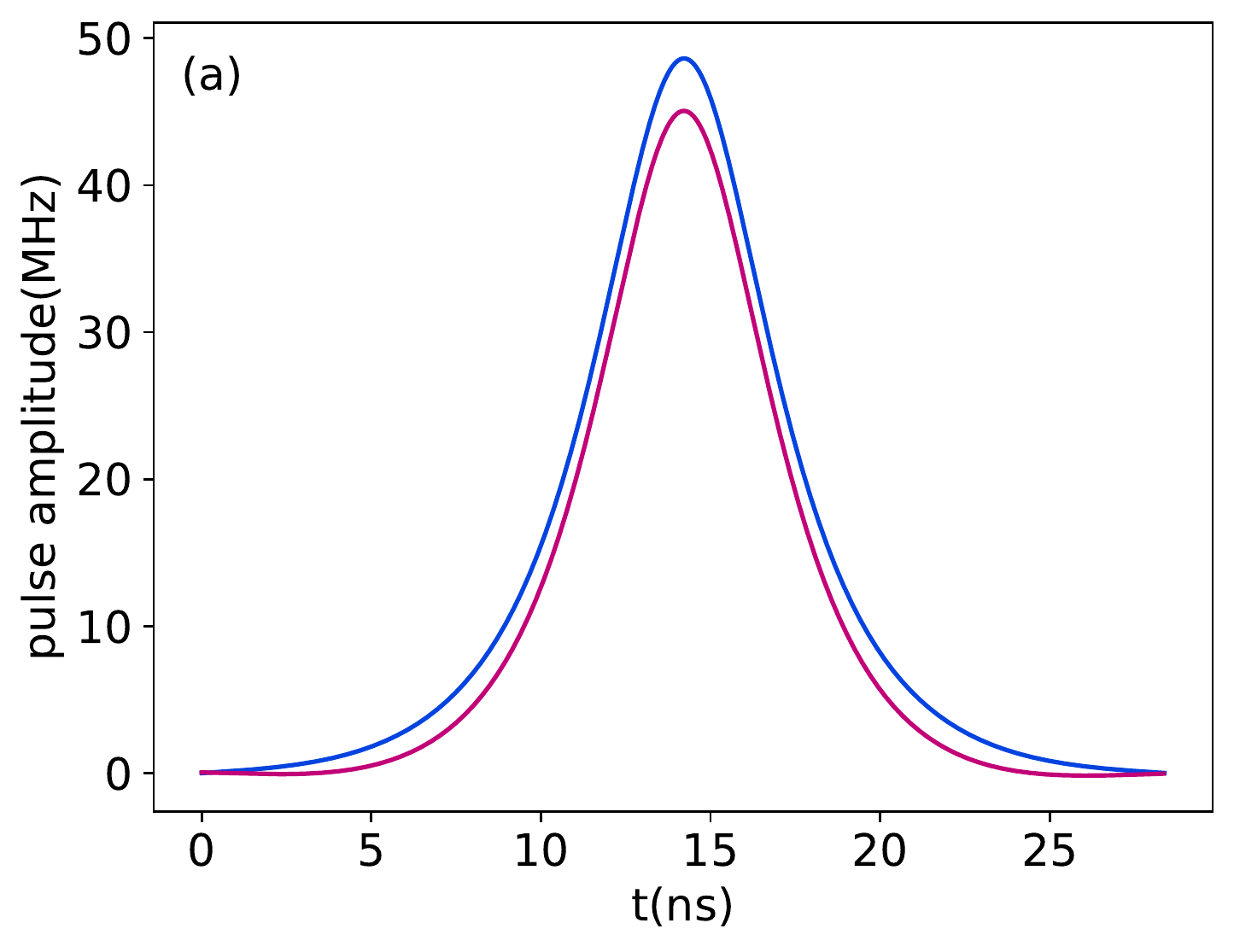}
    \end{subfigure}
    \hfill
    \begin{subfigure}
        \centering
        \includegraphics[width=0.45\textwidth]{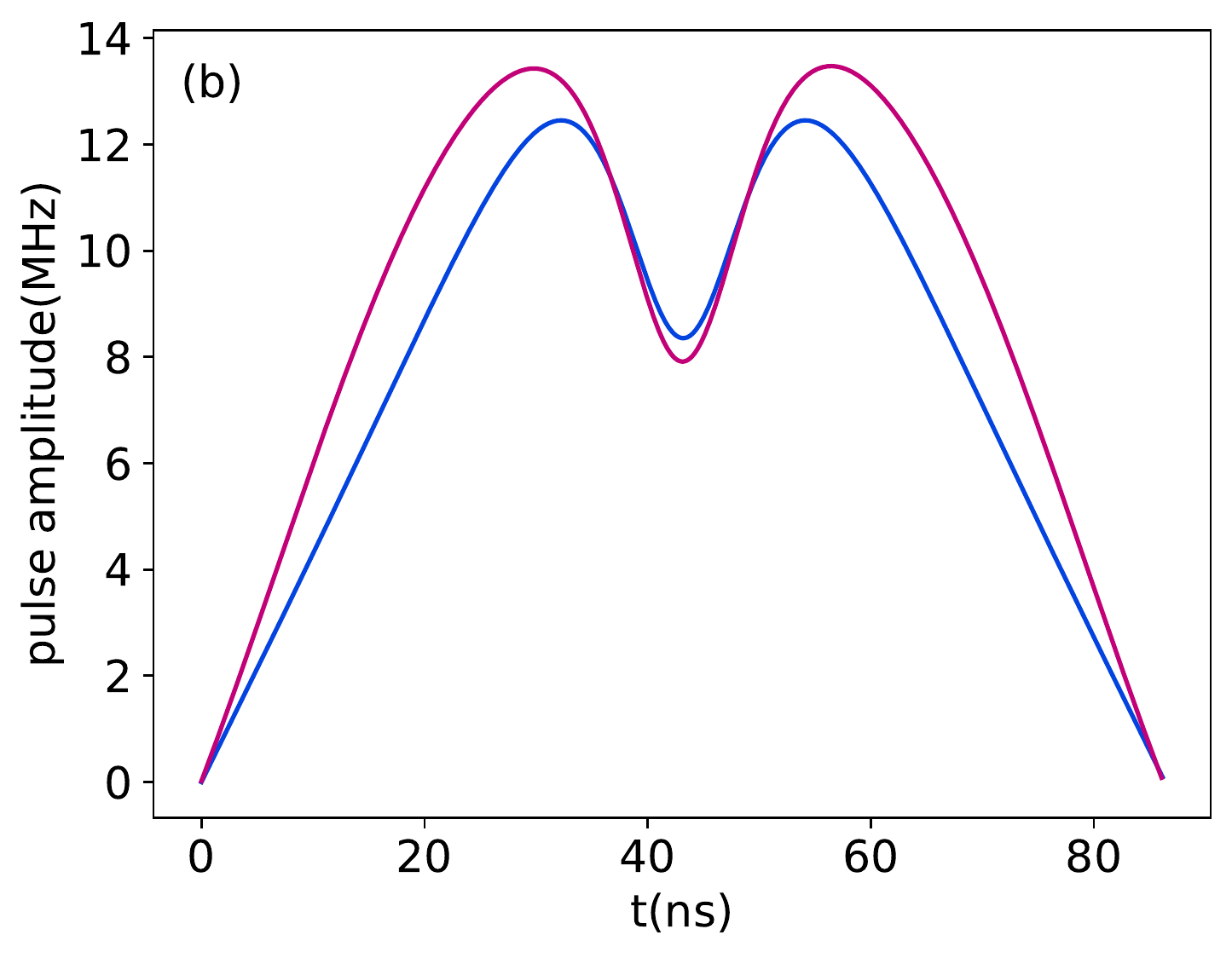}
    \end{subfigure}
    \caption{The initial geometric designed pulses (blue) shown in Fig.~\ref{fig:curve} and the GRAPE-optimized pulses (magenta). The optimized pulses are very close to the initial geometric pulses, but have infidelities five times lower.}
    \label{fig:opt_short_pulse}
\end{figure}

Fig.~\ref{fig:opt_short_pulse} shows the result of GRAPE optimization starting from the pulse shown in Fig.~\ref{fig:curve}. The optimized pulses, shown in magenta, are very close to the original pulses, and still quite smooth, but the infidelities after optimization have dropped by about a factor of $5$: from $2.51 \times 10^{-3}$ to $5.37 \times 10^{-4}$ for the short pulse (Fig.~\ref{fig:opt_short_pulse}(a)), and from $6.82 \times 10^{-3}$ to $1.65 \times 10^{-3}$ for the long pulse (Fig.~\ref{fig:opt_short_pulse}(b)). 

\begin{figure}
     \centering
     \begin{subfigure}
         \centering
         \includegraphics[width=0.45\textwidth]{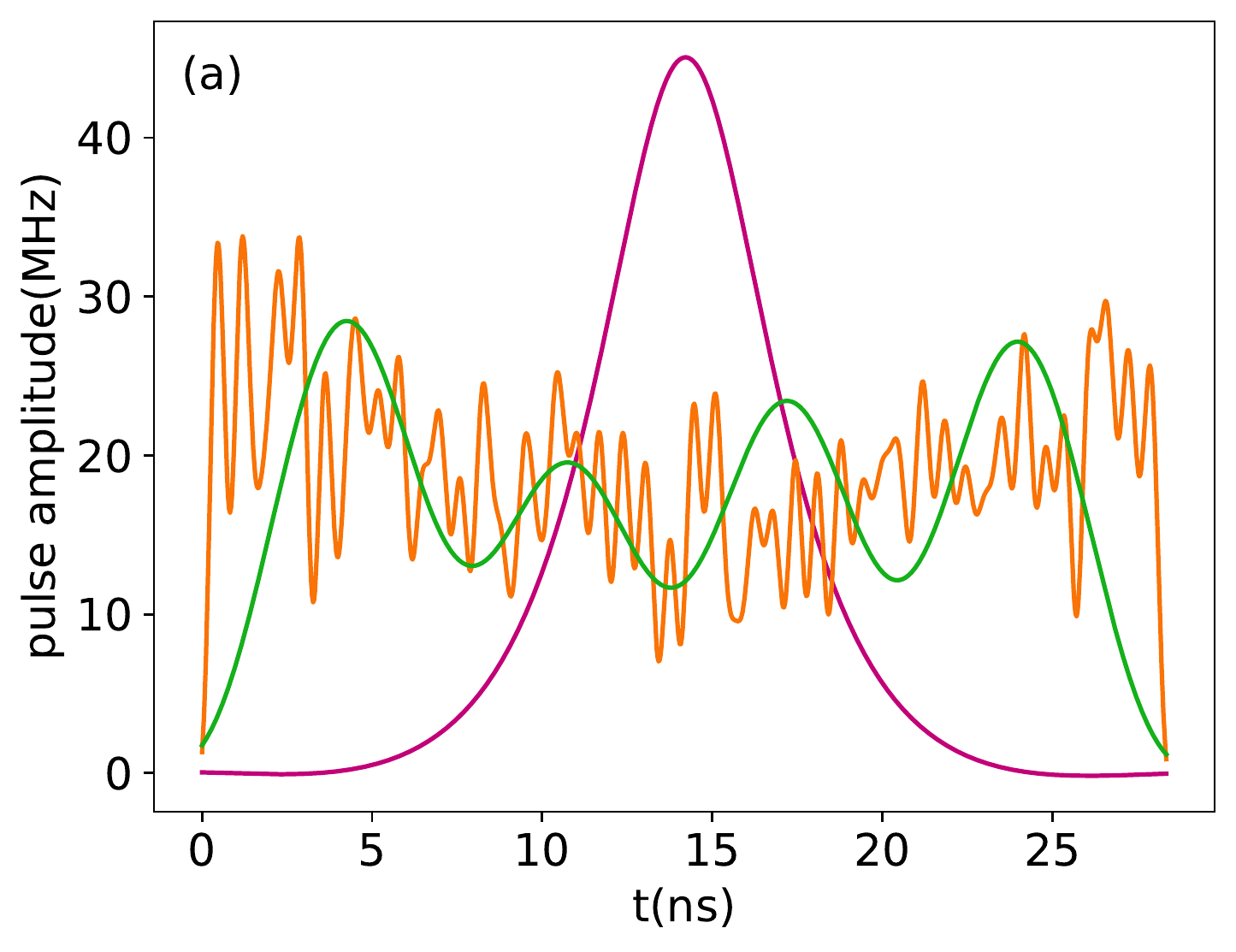}
     \end{subfigure}
     \hfill
     \begin{subfigure}
         \centering
         \includegraphics[width=0.45\textwidth]{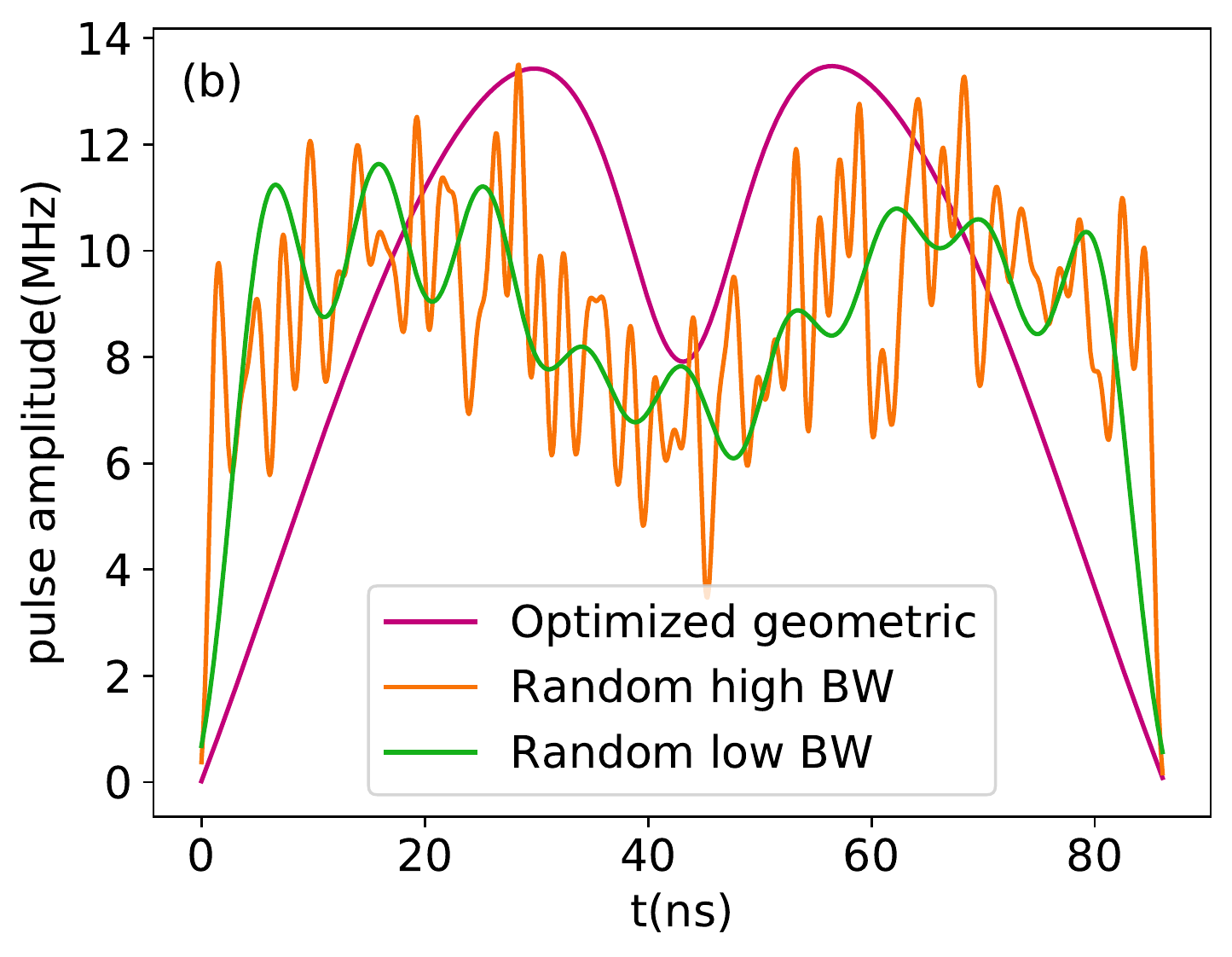}
     \end{subfigure}
        \caption{Here we compare the pulses obtained from GRAPE starting from a geometric pulse (magenta), and random Slepian pulses with high (orange) and low (green) bandwidth. Starting from a geometric pulse allows us to obtain simple pulses that outperform those obtained from Slepian sequences.}
        \label{fig:comparison}
\end{figure}

We compare the control pulses obtained in this way to those from the method introduced in Ref.~\cite{lucarelli2018quantum} for obtaining low-bandwidth control pulses, where the pulse is represented as a linear combination of bandwidth-constrained Slepian sequences \cite{slepian1978prolate}. Fig.~\ref{fig:comparison} shows a comparison of pulses obtained starting from a geometric pulse, a high-bandwidth Slepian pulse, and a low-bandwidth Slepian pulse. For the $28.4$ns pulses (Fig.~\ref{fig:comparison}(a)), the geometric pulse gives an infidelity of $5.37 \times 10^{-4}$, while the high- and low-bandwidth Slepian pulses only achieve an infidelity of $3.53 \times 10^{-3}$ and $4.32 \times 10^{-3}$, respectively. For the $86.2$ns pulses (Fig.~\ref{fig:comparison}(b)), the geometric pulse gives an infidelity of $1.65 \times 10^{-3}$, while the high- and low-bandwidth Slepian pulses achieve infidelities of $3.31 \times 10^{-3}$ and $1.59 \times 10^{-3}$, respectively. Thus we see that the optimized geometric pulses perform similarly to (if not better than) pulses obtained from Slepian sequences while yielding much simpler pulse shapes.

\section{Conclusion}
We introduced a geometrical approach for designing entangling gates that provides a global view of the optimal control landscape by mapping entanglement growth to geometric space curves. We illustrated the method by designing high-fidelity maximally entangling gates for silicon quantum dot spin qubits. 
We derived the minimal constraints on the space curves needed to guarantee the resulting gates have the desired entangling power. The pulses extracted from the geometric properties of these curves  
are smooth, have low bandwidth, and start and end at zero amplitude by design, making them experimentally feasible.
We showed that these pulses can be further improved by GRAPE optimization to obtain higher-fidelity operations while keeping the nice properties of the pulses.
Our work illustrates how the performance of numerical pulse optimization techniques can be further enhanced by exploiting global information about the optimal control landscape afforded by the geometrical perspective.

\section*{Acknowledgement}
This work is supported by the Army Research Office (grant nos. W911NF-15-1-0149 and W911NF-17-0287).

%


\end{document}